\documentclass[aps,prb,a4paper,twocolumn,groupedaddress,english,eqsecnum,showpacs,longbibliography]{revtex4-1}
\usepackage{babel}
\usepackage{xcolor}
\usepackage{graphicx}
\usepackage{amsmath}
\usepackage{amssymb}

\definecolor{darkblue}{rgb}{0.1,0.2,0.6}
\definecolor{darkred}{rgb}{0.8,0.1,0.2}
\usepackage[colorlinks,citecolor=darkblue,linkcolor=darkred,urlcolor=darkblue]{hyperref} 
\usepackage[all]{hypcap} 

\makeatletter
\adddialect\l@English\l@english
\makeatother

\def\be{\begin{equation}} \def\ee{\end{equation}}

\def\bea{\begin{eqnarray}} \def\eea{\end{eqnarray}}

\newcommand{\cf}{\textit{cf.} }
\newcommand{\ie}{\textit{i.e.} }

\begin{document}
\title{Universal logarithmic corrections to entanglement entropies
in two dimensions with spontaneously broken continuous symmetries}
\author{David J. Luitz}
\affiliation{Laboratoire de Physique Th\'eorique, IRSAMC, Universit\'e de Toulouse,
{CNRS, 31062 Toulouse, France}}
\author{Xavier Plat}
\affiliation{Laboratoire de Physique Th\'eorique, IRSAMC, Universit\'e de Toulouse,
{CNRS, 31062 Toulouse, France}}
\author{Fabien Alet}
\affiliation{Laboratoire de Physique Th\'eorique, IRSAMC, Universit\'e de Toulouse,
{CNRS, 31062 Toulouse, France}}
\author{Nicolas Laflorencie}
\affiliation{Laboratoire de Physique Th\'eorique, IRSAMC, Universit\'e de Toulouse,
{CNRS, 31062 Toulouse, France}}
\date{\today}

\begin{abstract}
    We explore the R\'enyi entanglement entropies of a one-dimensional (line) subsystem of length $L$ embedded in two-dimensional $L\times L$ square lattice for quantum spin models whose ground-state breaks a continuous symmetry in the thermodynamic limit. 
    Using quantum Monte Carlo simulations, we first study the $J_1 -
    J_2$ Heisenberg model with antiferromagnetic nearest-neighbor $J_1>0$ and ferromagnetic second-neighbor couplings $J_2\le 0$. The signature of SU(2) symmetry breaking on finite size
    systems, ranging from $L=4$ up to $L=40$ clearly appears as a universal additive logarithmic
correction to the R\'enyi entanglement entropies: $l_q \ln L$ with $l_q\simeq 1$, independent of the
R\'enyi index and values of $J_2$.  We confirm this result 
using a high precision spin-wave analysis (with restored spin rotational symmetry) on finite
lattices up to $10^5\times 10^5$ sites, allowing to explore further non-universal finite size
corrections and study in addition the case of U(1) symmetry breaking. 
Our results fully agree with the prediction $l_q=n_G/2$ where $n_G$ is the number of
Goldstone modes, by Metlitski and Grover [arXiv:1112.5166].  \end{abstract}
\maketitle

\section{Introduction} Entanglement entropy (EE) is now well recognized as a very powerful tool to
diagnose various quantum states of matter~\cite{Calabrese_Special_JPA09,grover_entanglement_2013}.
For interacting quantum systems in dimension $D\ge 2$, the ground-state EE of a given spatial partition $A$
embedded in a larger system scales with the perimeter $\ell_A$ of $A$, following the so-called
area-law~\cite{srednicki_entropy_1993,eisert_colloquium:_2010} for any R\'enyi index $q>0$ \be
S_q=\frac{1}{1-q}\ln\Bigl({\rm{Tr}}\left[{\hat{\rho}}_A\right]^q\Bigr)=a_q\ell_A+\cdots \label{EE}
\ee where ${\hat{\rho}}_A$ is the reduced density matrix of the subsystem $A$. While the leading
part $a_q\ell_A$ is not expected to reflect the universality of the phase, sub-leading terms (the
ellipsis in Eq.~\eqref{EE} above) may encode it, as first discovered for topological
order~\cite{kitaev_topological_2006,levin_detecting_2006}. For systems which exhibit a true
long-range order in the ground-state with a continuous symmetry breaking in the thermodynamic limit, the EE of a subsystem
has been predicted~\cite{metlitski_entanglement_2011} to exhibit a universal additive logarithmic
correction to the area law term, with a prefactor in dimension $D=2$ controlled by the number of
Goldstone modes $n_G$ associated to the broken symmetry: \be S_q=a_q\ell_A +
\frac{n_G}{2}\ln\ell_A+\cdots \label{eq:MG} \ee In such finite systems (with $N=L\times L$ sites),
there are two types of excitations: the Anderson tower of states (TOS)~\cite{anderson_approximate_1952} with an energy
scaling as $1/L^2$ and $n_G$ Goldstone modes (SW for quantum magnets) with a linear dispersion $\sim
1/L$, both being key contributions for the expected logarithmic corrections in
Eq.~\eqref{eq:MG}~\cite{metlitski_entanglement_2011}.  As first detected using a modified spin-wave
(SW) approach for the SU(2) symmetric Heisenberg antiferromagnet on a square
lattice~\cite{song_entanglement_2011}, subsequent quantum Monte Carlo (QMC)
calculations~\cite{kallin_anomalies_2011,humeniuk_quantum_2012,helmes_entanglement_2014} have also been able to capture
additive logarithmic corrections, while estimates of the prefactor did not agree with the
prediction. 
Also, the expected factor of two between the logarithmic terms for SU(2) and
U(1) was not clearly observed~\cite{humeniuk_quantum_2012}.
The possible reasons for such discrepancies are temperature and/or statistical effects, as
well as the importance of further finite-size corrections (beyond the log term) which might be hard to capture with QMC simulations on finite-size systems. Also one has to carefully subtract contributions to
logarithmic corrections to $S_q$ from corners if present in the subsytem geometry~\cite{kallin_anomalies_2011,helmes_entanglement_2014}: these
contributions are usually numerically small and their estimates from QMC simulations are suffering
from the above mentioned difficulties.
    \begin{figure}[t] \centering \includegraphics[width=0.485\columnwidth,clip]{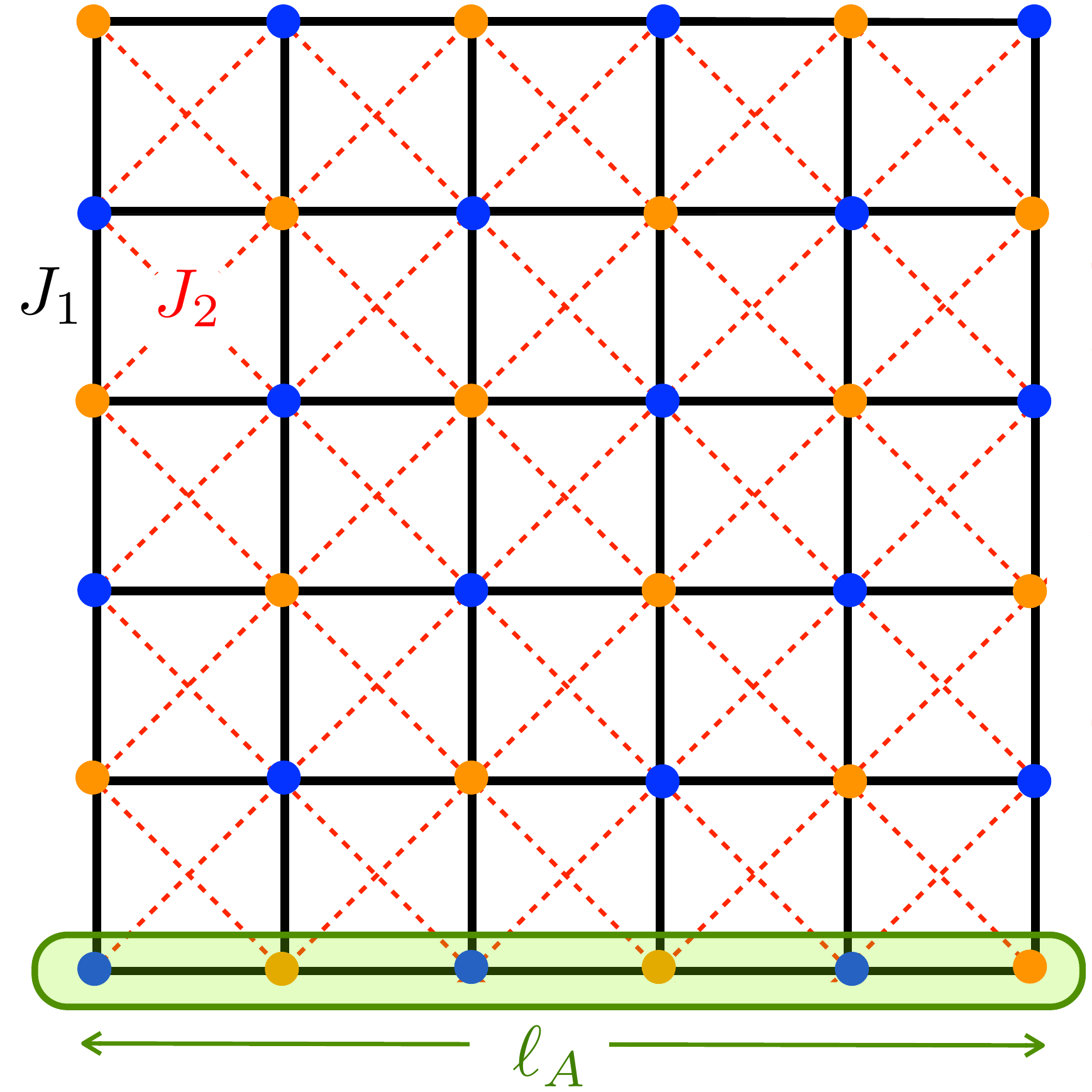}
    \caption{Schematic picture for the $J_1 - J_2$ square lattice. The line shaped subsystem $A$ of
length $\ell_A$ is shown in green.} \label{fig:lattice} \end{figure}
 \begin{figure*}[t] \includegraphics[width=\columnwidth]{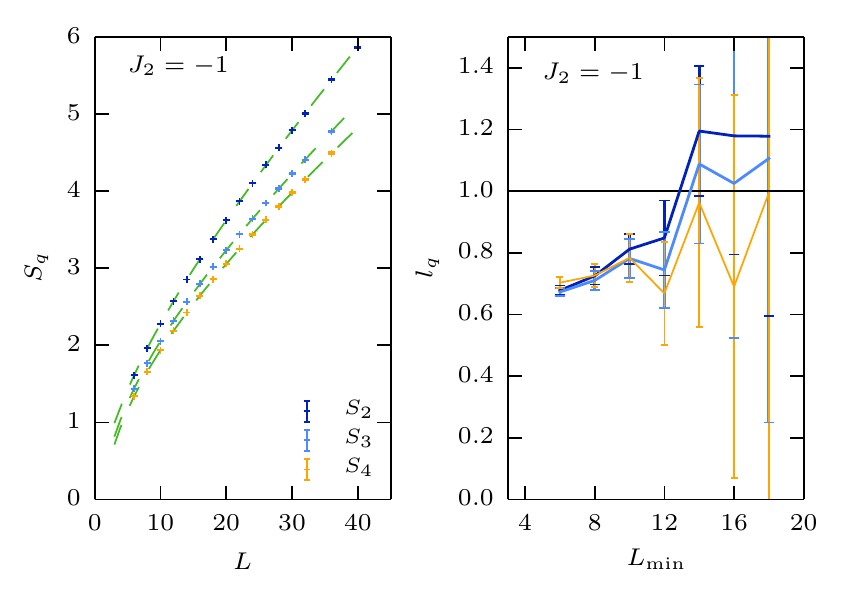} \hfill
     \includegraphics[width=\columnwidth]{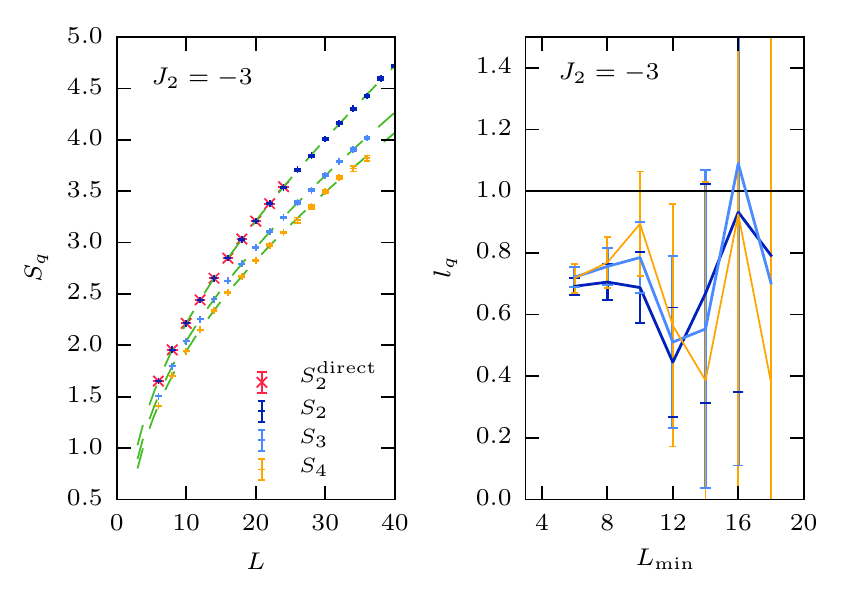} \caption{(Color online) QMC results for the
         entanglement R\'enyi entropies of the $J_1-J_2$ Heisenberg model for $J_2=-1$ (left two panels) and $J_2=-3$ (right two
         panels). We show the prefactor of the logarithmic scaling term obtained by fits to the form
 $S_q=a_q L + l_q \ln L + b_q + c_q/L$ over fit ranges $[L_\text{min},L_\text{max}]$ as a function of
 $L_\text{min}$, with $L_\text{max}=40$ for $q=2$ and $L_\text{max}=36$ for $q=3,4$. Our results are consistent with $l_q=1$ independent of $J_2$ and $q$. For $J_2=-3$, we also show
 the EE $S_2^\text{direct}$ obtained by a direct mixed ensemble calculation
 using the method of Ref. \onlinecite{humeniuk_quantum_2012}.}
 \label{fig:qmc} \end{figure*}

Very recently, Kulchytskyy {\it {et al.}} used an improved estimator for $S_2$ (for a half-torus
subsystem) with QMC simulations of the spin-$\frac{1}{2}$ XY model on the square
lattice~\cite{kulchytskyy_detecting_2015} which allowed them to get a quite precise estimate for the
prefactor of the log correction $\simeq 0.5$, fully consistent with $n_G=1$ Goldstone boson
associated to the breaking of U(1) symmetry. Nevertheless, to the best of our knowledge there is no
numerical study demonstrating the universality of Eq.~\eqref{eq:MG}, such as its independence on the
R\'enyi index $q$, details of microscopic Hamiltonian or type of continuous symmetry breaking.  

In this paper, we aim at going further to test the prediction Eq.~\eqref{eq:MG} for various values
of $q$ and for two quantum spin models having different symmetries, using a one-dimensional ring of
length $\ell_A=L$ as subsystem $A$ (see Fig.~\ref{fig:lattice}) embedded in a $L\times L$ torus.
This is the simplest possible corner-free bipartition scaling with $L$ where the universal
logarithmic correction proportional to the number of Goldstone modes should be present. 

We explore the R\'enyi EEs $S_q$ for such a subsystem using two techniques: exact
QMC simulations of $S_q$ with $q=2,3,4$ (Sec.~\ref{sec:QMC}), and a semi-classical SW theory for
finite size systems~\cite{takahashi_modified_1989,hirsch_spin-wave_1989} where spin rotational
invariance is restored such that both TOS and Goldstone modes are included (Sec.~\ref{sec:SW}). The
choice of a line subsystem is advantageous for these two techniques: in QMC, we use the improved
estimator introduced in Ref.~\onlinecite{luitz_improving_2014}, which is particularly efficient when
subsystem volume is as small as possible (it is in fact minimal for the line subsystem), while the SW calculations are particularly simplified by full translation symmetry of the
line subsystem, allowing for an analytical understanding of the $\frac{n_G}{2}\ln L$ term of Eq.~\eqref{eq:MG}.

\section{Quantum Monte Carlo results} \label{sec:QMC} 

For our quantum Monte Carlo calculations, we consider the spin-$1/2$ $J_1 - J_2$ antiferromagnet defined on a
bipartite $L\times L$ square lattice by the following Hamiltonian
\be {\cal H}_{J_1-J_2}=J_1\sum_{\langle i j\rangle}{\vec{S}}_i\cdot {\vec{S}}_j + J_2\sum_{\langle\langle i
j\rangle\rangle}{\vec{S}}_i\cdot {\vec{S}}_j, \label{eq:J1J2} \ee
where ${\vec{S}}$ are spin-$\frac{1}{2}$ operators, interactions act between nearest neighbors $\langle
ij\rangle$ and second nearest neighbours ${\langle\langle i j\rangle\rangle}$ along the diagonals of
the square lattice (see Fig.~\ref{fig:lattice}).  For this work, we consider antiferromagnetic nearest neighbors
interactions $J_1>0$ and {\it ferromagnetic} second neighbors interactions $J_2<0$, for which it is known that
the ground-state exhibits antiferromagnetic long-range order, thus breaking SU(2) symmetry
associated with two Goldstone modes (independent of $J_2<0$). The
motivation for adding the second neighbors interaction $J_2$ is to check the universality of the results with
respect to microscopical variations of the Hamiltonian (different values of $J_2$) without changing
the nature of the ground-state and of the low-lying excitations. Additionally, as $|J_2|$ is increased,
the antiferromagnetic long-range order is enhanced ({\it{i.e.}} larger values of the order parameter),
and we therefore expect lower EEs as we get closer to a classical Heisenberg
antiferromagnet.

  We perform extensive QMC simulations of this model for two different values of $J_2$ using the
  stochastic series expansion (SSE) algorithm~\cite{sandvik_quantum_1991,syljuaasen_quantum_2002}. We compute the R\'enyi EE $S_q$ for $q=2,3,4$ using
  a recently introduced decomposition\cite{luitz_improving_2014} that benefits from subsystem
  symmetries. This method is particularly useful when the surface of the subsystem scales as its
  volume, \ie if the subsystem volume is minimal without introducing geometrical effects, such as
  corners. In this sense, this method is optimal for the line shaped subsystem in Fig.~\ref{fig:lattice}.

  Our simulations are performed in the finite temperature formulation of the SSE at low enough
  temperatures in order to capture only ground-state physics. While the finite size gap of the tower of states
  scales as the inverse of the total number of spins $N=L\times L$ in the system, one would expect that it is necessary to scale the inverse temperature $\beta$ linearly in
  $N$. For the system sizes we studied (up to $L=40$), we find however that the results for the EE of simulations at inverse temperatures $\beta=8L$ and $\beta=4L$ agree within errorbars:
  we therefore performed all calculations at inverse temperatures higher than $\beta=4L$.
 Fig. \ref{fig:qmc} shows the QMC result of the line EEs as a function of system
 size for different R\'enyi indices and two values of $J_2$. For $J_2=-3$ and $q=2$, we also perform an independent set of QMC
  simulations in an extended ensemble where EE is directly computed from the ratio
  of partition functions \cite{humeniuk_quantum_2012}. We obtain a perfect agreement between the two methods. Note, that the decomposition of the EE as described in Ref. \onlinecite{luitz_improving_2014} allows us to access larger system sizes (in particular for higher R\'enyi indices) with very high precision.

We fit our results for the line EEs to the scaling
 ansatz
 \be
 S_q = a_q \ell + l_q \ln \ell + b_q + c_q/L
 \label{eq:scaling_ansatz}
 \ee
 to infer if the prefactor of the logarithmic term is indeed $l_q=n_G/2$ ($=1$ for the ground-state of the model Eq.~\ref{eq:J1J2}). We systematically
 reduce the fitting range $[L_\text{min},L_\text{max}]$ of included system sizes, always including
 the largest systems with $L_\text{max}\sim 40$ and studying the best fit value of $l_q$ as a
 function of $L_\text{min}$ as shown in the right panels of Fig. \ref{fig:qmc}. Note that the errorbars stem from a careful bootstrap study of the stability of the fit
 introducing gaussian resampling of the data and perturbations of the initial parameters. We have
 studied the statistical behavior of the fit-data distance quantified by $\chi^2$ and find that
 the qualities~\cite{young_everything_2012} $Q$ of the best fit are already very good (around $0.7 \dots
 0.9$) using the scaling ansatz from equation \eqref{eq:scaling_ansatz}. Hence, the quality of our
 data does not allow for an inclusion of higher order terms (which could result in overfitting
 statistical noise).

While the log term is clearly present as nicely visible in the concavity of $S_q(L)$, we found it
difficult to get a very precise estimate for the prefactor $l_q$ in the thermodynamic limit. What
is clear however is that whereas the area law term does depend on the R\'enyi index $q$ and $J_2$, there is
apparently no $q$-dependence for $l_q$. Taking into account the largest $L_\text{min}$, we can estimate that $l_q=1.0(3)$, fully compatible with the prediction $l_q=1$, albeit with admittedly large error bars. In order to reach much larger systems (which would be helpful in studying the convergence of $l_q$ with $L_\text{min}$), we now consider a SW calculation of EE for the same setup of a line subsystem.

\section{Spin-Wave theory} \label{sec:SW}

Modified SW theory for finite size
systems~\cite{takahashi_modified_1989,hirsch_spin-wave_1989} has been shown to be very useful for
computing EEs of the square lattice Heisenberg antiferromagnet in
Ref.~\onlinecite{song_entanglement_2011}. The crucial point is to artificially restore the spin rotational
invariance in order to mimic the symmetric ground-state of a finite size system. For this, a size-dependent
regularizing external field $h^*$ is imposed to the system such that the SW-corrected order
parameter is identically zero. 

We study the $J_1 - J_2$ Heisenberg antiferromagnet Eq.~\eqref{eq:J1J2} where SU(2) symmetry is
restored by adding a small staggered magnetic field
\be {\cal H}_{J_1-J_2}=J_1\sum_{\langle i j\rangle}{\vec{S}}_i\cdot {\vec{S}}_j + J_2\sum_{\langle\langle i
j\rangle\rangle}{\vec{S}}_i\cdot {\vec{S}}_j +h^*\sum_{i}(-1)^iS_i^z, \label{eq:J1J2h} \ee
such that $\langle S_i^z\rangle=0$, as well as the ferromagnetic XY model
\be {\cal H}_{{\rm XY}}=-J\sum_{\langle i j\rangle}\left(S_i^x S_j^x +S_i^yS_j^y\right)+h^*\sum_{i}S_i^x,
\label{eq:XYh} \ee
where the transverse field $h^*$ is chosen such that $\langle S_i^x\rangle=0$ {and $\langle
S_i^y\rangle = 0$} in order to
artificially restore the U(1) symmetry. In both cases, the external field $h^*\propto L^{-4}$ leads to a finite
size gap $\Delta\sim \sqrt{h^*}\sim 1/L^2$, below the $n_G$ linearly
dispersing Goldstone modes ($n_G=2$ for the SU(2) $J_1-J_2$ model, and $n_G=1$ for the U(1) XY
model). This additional energy scale reproduces the TOS structure on finite systems. For the analytical expressions below, we do not specify the value of the spin $S$, while for the numerical computations we explicitly consider $S=1/2$.

The calculation of the EEs in the SW approximation is eased by the quadratic
nature of the SW Hamiltonian (at the linear harmonic level), as the reduced density matrix can be expressed as an exponential of a
correlation matrix $C$ involving only expectation values of two-point correlation
functions~\cite{plenio_entropy_2005,barthel_entanglement_2006}. The EEs of a subsystem composed of $N_A$
sites are obtained as a function of the $N_A$ eigenvalues $\nu_p^2$ of this correlation matrix: \be
{S}_q =\frac{1}{q-1} \sum_p \ln\left[\left(\nu_p+\frac{1}{2}\right)^q-
\left(\nu_p-\frac{1}{2}\right)^q\right], \label{eq:Sq}\ee and for $q=1$: \bea
S_1=\sum_p&\Bigl[&\left(\nu_p+\frac{1}{2}\right)\ln\left(\nu_p+\frac{1}{2}\right)\nonumber\\
          &-&\left(\nu_p-\frac{1}{2}\right)\ln\left(\nu_p-\frac{1}{2}\right)\Bigr].  \eea
The eigenvalues $\nu_p^2$ are typically obtained from numerical diagonalization of the matrix $C$,
once its elements have been evaluated for the value $h^*$ (which depends on system size and
parameters such as $J_2$ or the spin value $S$). 
In the case of a line subsystem, the numerical diagonalization step can be circumvented by
noticing that the translation invariance along the line implies that the
matrix $C$ is \emph{circulant}\footnote{For the SU(2) case more precisely, the matrix $C$ is block-diagonal with two identical circulant blocks (corresponding to correlations between odd and even sublattice sites of the bipartite lattice), and each block contributes half of the eigenvalues mentioned in Eq.~\eqref{eq:nu}.}: its eigenvalues are given by the Fourier transform of its first
line.
Plugging in all the expectation values and exploiting the convolution theorem, we obtain 
\be \nu_p = \frac{1}{2L}\sqrt{
\left(\sum_{k_y} \frac{A{(p,k_y)}}{\Omega{(p,k_y)}}\right)^2 - \left(\sum_{k_y}
\frac{B{(p,k_y)}}{\Omega{(p,k_y)}}\right)^2}, 
\label{eq:nu}
\ee 
where $p=-\pi+\frac{2\pi}{L}(j-1)$ with $j \in [1,L]$, and the SW excitation spectrum $\Omega{(\bf k)}=\sqrt{A({\bf k})^2 -B({\bf k})^2}$ is given by
\bea
A(k_x,k_y)&=&2SJ_2\cos k_x\cos k_y+ 2S(J_1-J_2)+\frac{h^*}{2}\nonumber\\
B(k_x,k_y)&=&-SJ_1\left[\cos k_x + \cos k_y\right]
\label{eq:ABSU2}
\eea
for the $J_1-J_2$ model, while for the XY case
\bea
A(k_x,k_y)&=&-S\frac{J}{4}\left[\cos k_x + \cos k_y\right]
+SJ+\frac{h^*}{2}\nonumber\\
B(k_x,k_y)&=&S\frac{J}{4}\left[\cos k_x + \cos k_y\right].
\label{eq:ABU1}
\eea
The symmetry of the line subsystem allows to directly compute EEs within the SW approximation using Eq.~\eqref{eq:nu} for systems of
very large linear size (up to $L\sim 10^5$), which can hardly be reached if a numerical diagonalization of
$C$ is involved.  Note that for such large systems, the regularizing field $h^*$ becomes extremely
small, and we resorted to arbitrary-precision numerics to ensure convergence of $h^*$ and
corresponding $\nu_p$.

\begin{figure}[b] \centering \includegraphics{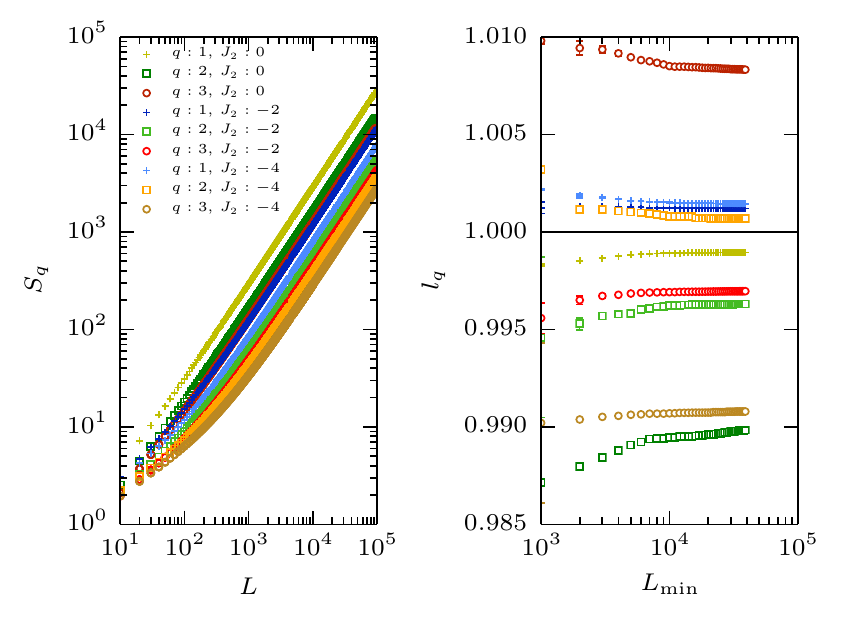} \caption{(Color online) Left panel:
        EE $S_q$ of the line shaped subsystem in the $J_1-J_2$ model for
        different values of $J_2$ and R\'enyi indices $q$ as obtained from the modified spinwave
        analysis. Right panel: Prefactors $l_q$ of the logarithmic corrections obtained
from fits of the form $albl'l''c'$ (\cf Eq. \eqref{eq:fitfunctions} for definitions of the terms) as
a function of the minimal size $L_\text{min}$ included in the fit.} \label{fig:EE_sw_xxx}
\end{figure}
\begin{figure} \centering \includegraphics{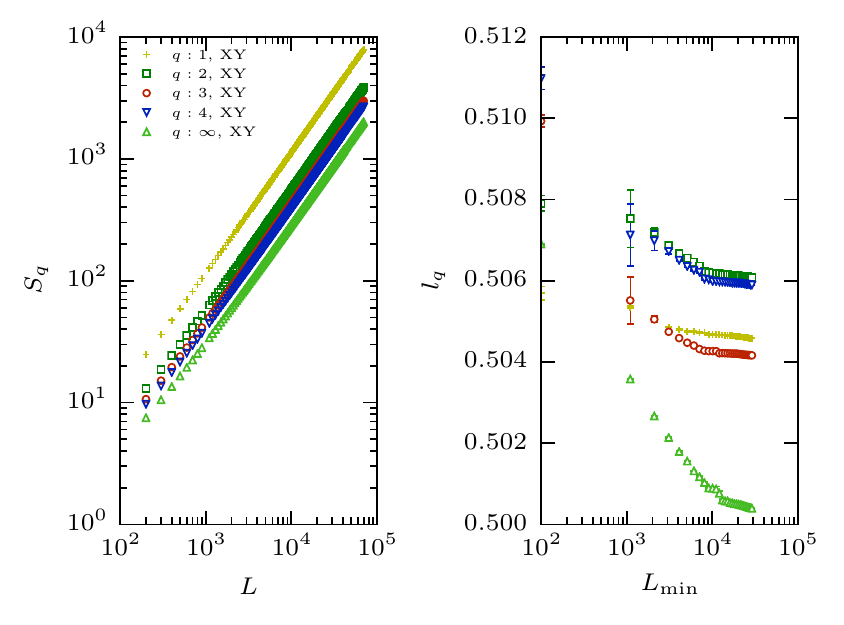} \caption{(Color online) Left panel:
        EE $S_q$ of the line shaped subsystem in the XY model for
        different R\'enyi indices $q$. Right panel:  Logarithmic term in
        the scaling of the EE of a line in the XY model for different fit ranges
        $[L_\text{min},7\cdot 10^4]$ and different R\'enyi indices $q$, as obtained from a fit of the form $albl'l''c'c$ (see Eq. \eqref{eq:fitfunctions}).} \label{fig:EE_sw_xy}
\end{figure}

\begin{figure}[h] 
    \centering 
    \includegraphics{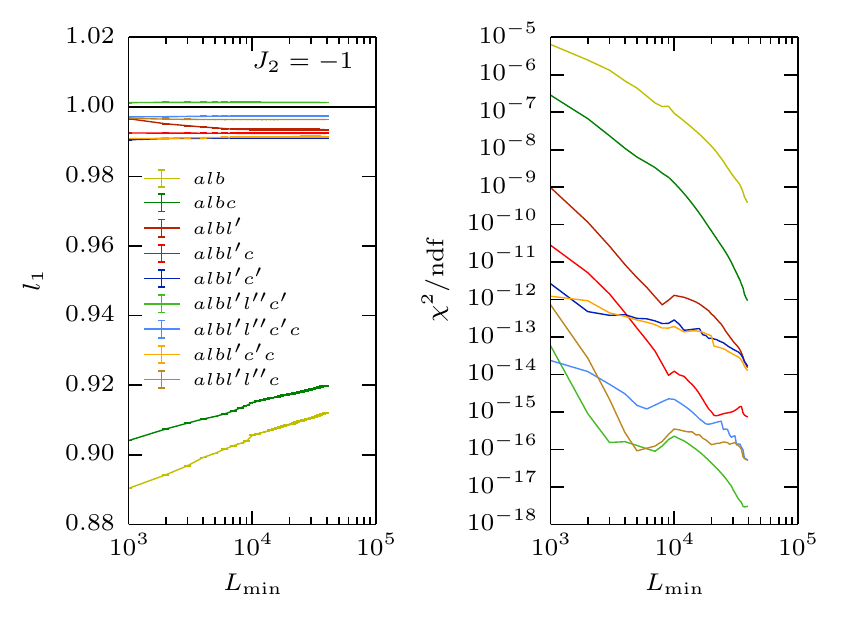}
    \caption{(Color online) Comparison of different fits over the range $[L_\text{min},10^5]$ to the spin wave
        result for $S_1$ at $J_2=-1$ in the $J_1-J_2$ Heisenberg model. The left panel displays the prefactor $l_1$
        of the logarithmic scaling term $l_1 \ln(L)$, which all fits find to be very close to unity.
        The right panel displays the corresponding $\chi^2$ normalized by the number of degrees of
        freedom (ndf). Clearly the best fits with the lowest $\chi^2$ find $l_1$ to be closest to
        $1$. The artifacts around $L_\text{min}\approx 10^4$ stem from a change of the grid on which
we calculated $S_1$ which effectively introduces a higher weight for points in the denser region of
the grid at smaller system size. The fit functions are coded according to the terms in equation
\eqref{eq:fitfunctions}.  } 
\label{fig:xxx_sw_fitcomp} 
\end{figure}

\begin{figure}[h] 
    \centering 
    \includegraphics{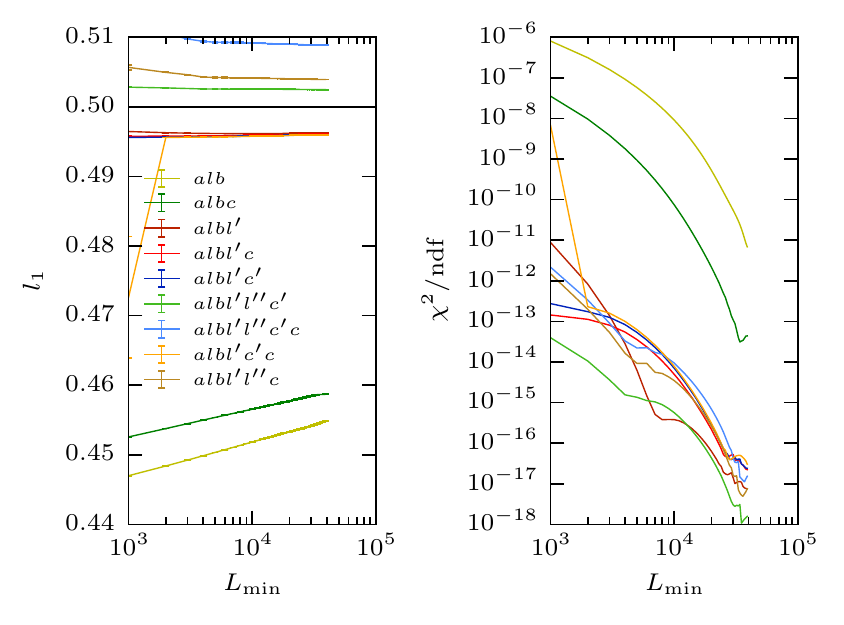}
    \caption{(Color online) Comparison of different fits over the range $[L_\text{min},7\cdot 10^4]$ to the spin wave
        result for $S_1$ in the $XY$ model. The left panel displays the prefactor $l_1$
        of the logarithmic scaling term $l_1 \ln(L)$, which all fits find to be very close to one
        half. The right panel displays the corresponding $\chi^2$ normalized by the number of degrees of
        freedom (ndf). Clearly the best fits with the lowest $\chi^2$ find $l_1$ to be closest to
        $0.5$. The fit functions are coded according to the terms in equation
\eqref{eq:fitfunctions}.  } 
\label{fig:xx_sw_fitcomp} 
\end{figure}

Considering now the case $S=1/2$, our numerical results for $S_q$ for different R\'enyi indices $q$ are displayed in Fig. \ref{fig:EE_sw_xxx} (left) for the $J_1-J_2$ Heisenberg model (for different values of the diagonal coupling $J_2$) and in
Fig.
\ref{fig:EE_sw_xy} (left) for the XY model. The precise value of $S_q$ being dominated by a non-universal area law term, one cannot directly compare the actual estimates of $S_q$ obtained within SW to our exact QMC results at small sizes due to the approximations inherent to the SW approach. However, we expect the universal subleading logarithmic scaling term to be well captured by the modified SW theory. Indeed, fitting our SW data to the previous form Eq.~\eqref{eq:scaling_ansatz} clearly yields an additive logarithmic term, as shown in Fig.~\ref{fig:xxx_sw_fitcomp} for the $J_1 -J_2$ antiferromagnet with $J_2=-1$ and in Fig.~\ref{fig:xx_sw_fitcomp} for the XY model. 
The slow convergence of the coefficient of the logarithmic term suggests that subleading corrections beyond the log term in
Eq.~\eqref{eq:scaling_ansatz} have to be included. As we are not aware of any prediction for such subleading corrections, we perform fits using the general ansatz 
\bea
S_q &=& a_q L + l_q \ln L + b_q \nonumber\\
&+&
l^2_q \ln \ln L + l^3_q \ln \ln \ln L + \frac{c_q}{L}+c^1_q \frac{\ln L}{ L},
\label{eq:fitfunctions}
\eea
leaving out systematically various terms. We use the
shorthand notation $albl^2l^3c^1c$ to label the various fit functions in the following figures
(terms whose parameters do not appear in this string are not included in the fits).  We find
nonvanishing contributions for all terms and comparing carefully the distance of the fit to the data
quantified by $\chi^2$, it seems that the inclusion of all these terms yields the best fits. We show
a representative analysis of different fit functions in Fig. \ref{fig:xxx_sw_fitcomp} for the
$J_1-J_2$ model and in Fig. \ref{fig:xx_sw_fitcomp} for the XY model. The comparison of the distances of the studied fit functions to the data shown in the right panels of Figs. \ref{fig:xxx_sw_fitcomp} and \ref{fig:xx_sw_fitcomp} indicates that the most reliable description of the data is obtained by the ansatz $albl^2l^3c^1$, which seems reasonable as the term $c^1 \ln L/L$ decreases
slowly and may therefore still be important at the available system sizes.

A word of caution is in order here regarding the meaning of $\chi^2$. This quantity is usually
normalized by (gaussian) statistical errorbars attached to the data and should therefore follow the
$\chi^2$ distribution. In particular, this implies that $\chi^2/\text{ndf}$ for a perfect fit
approaches unity and can not be smaller unless the model ``overfits'' statistical noise. Here, the
situation is strikingly different as our data do not bear statistical errorbars and $\chi^2$ does
not have any statistical meaning. In fact, for a perfect fit, $\chi^2$ would then vanish, a
situation we are very close to. The slow growth of the different fitting terms as
well as the fact that the exact form of the subleading terms in the scaling below the logarithmic
term remain unknown still gives rise to a small uncertainty of our fit results. 

Despite this, the different results for the investigated \emph{ans\"atze} consistently yield a
logarithmic prefactor which is very close to (or evolves with growing system sizes into) $l_q=1$ for
the  $J_1-J_2$ Heisenberg model and $l_q=1/2$ for the XY model. This can be clearly seen in Figs. \ref{fig:xxx_sw_fitcomp} and \ref{fig:xx_sw_fitcomp} for $l_1$, and for $l_q$ in Figs.~\ref{fig:EE_sw_xxx} and \ref{fig:EE_sw_xy} for different values of $q$ (as well as different $J_2$ for the $J_1-J_2$ Heisenberg model).

Our high-precision spin-wave results for a line subsystem are therefore in full agreement with the prediction Eq.~\eqref{eq:MG} of a prefactor $l_q=n_G/2$ reflecting the number of Goldstone modes associated with the breaking of a continuous
symmetry.

From the structure of the eigenvalues of the correlation matrix Eq.~\eqref{eq:nu} one can go further to interpret the additive logarithmic term in terms of the number of Goldstone modes $n_G$. Indeed, one can rewrite them as
\be
\nu_p=\frac{1}{2L}\sqrt{\sum_{k_y}\Theta(p,k_y)\sum_{k_y}\Theta^{-1}(p,k_y)},
\ee
with
\be
\Theta(p,k_y)=\sqrt{\frac{A(p,k_y)-B(p,k_y)}{A(p,k_y)+B(p,k_y)}},\,
\ee
$A$ and $B$ being given by Eqs.~\eqref{eq:ABSU2} and \eqref{eq:ABU1}, and the $L$ modes $p=-\pi+\frac{2\pi}{L}(j-1)$ with $j=1,\ldots,L$. It is straightforward to see that all $\Theta$ are non-singular $O(1)$ numbers, except at the singular points where Goldstone modes vanish. More precisely for SU(2) there are two contributions \be \Theta^{\rm {SU(2)}}(0,0)=\frac{1}{\Theta^{\rm{SU(2)}}(\pi,\pi)}\simeq\sqrt{\frac{8SJ_1}{h^*}}\simeq 2Nm_{\rm AF},\ee
where $m_{\rm AF}$ is the thermodynamic limit (SU(2) broken) staggered magnetization, and one contribution for U(1)
\be \Theta^{\rm {U(1)}}(0,0)\simeq\sqrt{\frac{SJ}{h^*}}\simeq 4Nm_{\rm xy},\ee
where $m_{\rm xy}$ is the transverse order in the thermodynamic limit.
Therefore all eigenvalues $\nu_p$ are $O(1)$ away from the Goldstone points where instead
\be
\nu_{\rm Goldstone}\propto \sqrt{L}+{\rm constant}.
\ee
Plugging this into the expression of the R\'enyi EEs Eq.~\eqref{eq:Sq}, the $L-n_G$ modes with $O(1)$ eigenvalues will add up and contribute $\sim L$ (the area law part) to $S_q$ and the $n_G$ terms will each contribute $\frac{1}{2}\ln L,~\forall q$. 
\section{Discussions and conclusions}

    We have investigated predictions from field theory that spontaneous breaking of a
    continuous symmetry leads to a logarithmic subleading scaling of the EEs $S_q$
    independent on microscopic parameters and the R\'enyi index $q$, in the specific case of a periodic line subsystem embedded in a two-dimensional torus. Our results, obtained using two different methods (numerically exact QMC and spin wave theory), are in perfect agreement with the prediction that the prefactor of the logarithmic term is given by $l_q=n_G/2$ by studying two models breaking SU(2) and U(1) symmetry respectively.

    Interestingly, we find that it is not necessary to study a bipartition of the system in two
    equal parts as cutting out a one dimensional subsystem is sufficient to capture the universal logarithmic correction. This is
    beneficial for both methods used in this work and we believe that other numerical and analytical
    techniques can profit from this finding in order to push calculations to larger system sizes, which
    are of tremendous importance for fitting the logarithmic term. Moreover, the spin-wave theory of the entanglement entropy of a line subsystem allows simplified calculations where the contribution of each Goldstone mode can be fully understood analytically in the modified (symmetry restored) spin-wave theory formalism.
    
    Reaching very large system sizes allowed us to capture higher order finite size corrections which demonstrates that it is very difficult to get a precise and size-converged estimate for the prefactor of the logarithmic correction $l_q$ using QMC simulations, restricted to linear sizes of a few tens of sites.
    
    Beyond this case of continuous symmetry-breaking phases, it would be interesting to
investigate whether the line subsystem can also capture
subdominant universal corrections associated with other
types of phases, such as discrete symmetry-breaking or topological phases. Indeed in the latter case, a one-dimensional geometry, as used in Ref.~\onlinecite{furukawa_topological_2007}, appears computationally more tractable (especially within QMC) than the usual topological entanglement entropy constructions~\cite{kitaev_topological_2006,levin_detecting_2006,isakov_topological_2011}. 

    \begin{acknowledgments}
    It is our pleasure to thank G. Misguich and M. Oshikawa for inspiring discussions and
    collaborations on related topics. 
    X.P. acknowledges Y. Fuji for interesting suggestions.This work was performed using HPC resources from GENCI (grant x2015050225) and CALMIP (grant 2015-P0677), and is supported by the French ANR program ANR-11-IS04-005-01. Our QMC simulations partly use the ALPS libraries~\cite{ALPS2}.
    \end{acknowledgments}

\bibliography{EE} 
\end{document}